\documentstyle[11pt,epsfig]{article}
\begin{document}
\title{\bf Enhanced roughness of lipid membranes caused by external electric fields}
\author{M Neek-Amal\thanks{neek@nano.ipm.ac.ir}$\,\,^1$, H.
Rafii-Tabar$^{1,2}$ and H. R. Sepangi$^{1,3}$\\
\footnotesize{$^1$Department of Nano-Science, Institute for
Studies in Theoretical Physics and
Mathematics (IPM)}\\ \footnotesize{ P.O. Box 19395- 5531,Tehran, Iran.}\\
\footnotesize{$^2$Research Center for Medical Nanotechnology and
Tissue Engineering}\\ \footnotesize{ Shahid Beheshti Medical
University, Evin, Tehran, Iran.}\\ \footnotesize{$^3$Department of
Physics, Shahid Beheshti University, Evin, Tehran, 19839, Iran}}
\maketitle
\begin{abstract}

The behavior of lipid membranes in the presence of an external
electric field is studied and used to examine the influence of such
fields on membrane parameters such as roughness and show that for a
micro sized membrane, roughness grows as the field increases. The
dependence of bending rigidity on the electric field is also studied
and an estimation of thickness of the accumulated charges around
lipid membranes in a free-salt solution is presented.\vspace{5mm}
\end{abstract}

\section{{  Introduction}}
Membranes can be modeled as statistical-mechanical systems, showing
a variety of configurations as $2D$ fluid-like surfaces. In addition
to thermal fluctuations, the fluctuation spectrum of a lipid
membrane can be influenced by several other external factors such as
osmotic pressure or pH differences. Another significant factor
contributing to these fluctuations is the presence of an external
electric field which can induce dramatic changes on the way a
membrane behaves, rupture or  shape transformation \cite{karin}
being one example. This is due to the fact that when an electric
field is applied to a lipid membrane it causes an extra lateral
tension (referred to as electrotension) to appear which depends on
the strength of the applied field as well as on a few constitutive
parameters of the lipid membrane \cite{Abidor,Needham,karin,hanc}.

Electrostatic interactions influence membrane elastic parameters in
a fundamental way, in particular the stability of flexible membranes
\cite{ywkim}. The non-vanishing excess membrane charge causes an
increase in membrane undulations due to Coulomb repulsion, while
charge fluctuations or free ions in a solution with screening
effects suppress this instability. It has been shown that
\cite{ywkim} with respect to free-ion or ion solvents, the surface
tension always decreases whereas bending rigidity may either
decrease or increase. Such instabilities are due to changes in
elastic parameters and may lead to a roughening of the lipid
membrane surface when exposed to thermal and electrical
fluctuations.

Roughening effects have important consequences on  mechanisms
governing the motion of molecules or particles along the
bio-membrane surface. Membrane roughness affects the interaction
between the surface and outer particles in one hand and the motion
of external inclusions \cite{langmuir,colloid} on the other. Also,
roughness fluctuations affect the electronic properties of a
membrane. These fluctuations are observed in graphene sheet and have
an important role in its electronic properties \cite{graphene}. In
this regard, the relevant parameter attributed to a membrane is the
roughness, defined as the root-mean-square value of vertical
variations \cite{langmuir} and depends on several membrane
parameters such as bending rigidity, surface tension, linear size,
temperature and so on \cite{encylopedia,Dietz}. For a detailed
discussion see \cite{colloid} and the references therein. Molecular
and monte carlo simulations have also been used to extract
Nano-Scale lipid membrane roughness using various kinds of
interaction parameters \cite{chemlett,jchemphys}. Apart from
roughness, other phenomenon such as electroporation in the case of
giant vesicles or a patch of membranes \cite{isam2}, lead to
important biotechnology applications.

Charge currents across a nearly flat and poorly conductive lipid
membrane give rise to out-of-equilibrium membrane undulations. In
this regard, two variant membranes under the influence of an
electric field, namely, a fluid membrane attached to a rigid frame
and a freely floating membrane can be considered. The theory behind
the above mentioned cases may be inferred from a general situation
where an infinite membrane is subjected to an external electric
field. In such a case, an effective negative surface tension appears
in the membrane, making it tense yet floppy-looking \cite{isam1}.
This theory is our preferred approach in computing and describing
the roughening and other quantities relating to lipid membranes. The
mean dynamical \emph{roughness} of the membrane can be obtained by
calculating the height-height correlation function \cite{Seifert1}.

In previous works \cite{Raf1,Raf2}, we computed the stochastic
trajectories of objects (inclusions) existing within and on the
surface of a membrane via the application of Langevin dynamics.
Also, by introducing the height function as a stochastic Wiener
process, a relation between random fluctuations of height and
lateral diffusion of membranes was studied \cite{Raf2}. In this
paper, we investigate the effects of electric fields on surface
tension and bending rigidity and their influence on the roughness in
lipid membranes. Our analysis is based on the extension of the
results previously presented in \cite{ywkim} and \cite{isam1} for
the bending rigidity and surface tension. For simplicity, we employ
an almost infinite flat lipid membrane subject to an electric field.
One result is that the weaker the electric field, the smaller the
roughness. If a slab made of charges with a certain thickness is
considered around the undulated membrane, the dependence of the
thickness on the electric field and other lipid membrane parameters
can be estimated.

The paper is organized as follows: in section two we write the
modified Canham-Helfrich Hamiltonian where the surface tension and
bending rigidity include a term resulting from the application of
the electric field. This Hamiltonian is then used to define the
dynamical roughness of the membrane in section three and conclusions
are drawn in the last section.

\section{Energetics of an almost plane membrane}
The free elastic energy of a symmetric, nearly flat, membrane is
described by the Canham-Helfrich Hamiltonian \cite{Can,Hel}. To
study a nearly flat membrane, it is convenient to consider it
parallel to the $(x_1,x_2)$ plane, regarded as the reference
plane. A single-valued {\it height} function $h$ represents the
position of a point on a fluctuating, nearly flat, sheet relative
to the reference plane and in this, the so-called {\it Monge
representation} \cite{Seifert1} the Hamiltonian is written as
\begin{eqnarray}
{\cal H}=\frac{1}{2}\int d^2x\left\{\kappa_0\left(\nabla^2 h
\right)^2+\sigma_0(\nabla h)^2\right\}, \label{eq1}
\end{eqnarray}
where $\sigma_0$ is the surface tension and $\kappa_0$ is the
bending rigidity of the membrane. In this form, the Hamiltonian is
expressed solely in terms of the height function, $h$, and its
derivatives. The Hamiltonian in (\ref{eq1}) describes the
energetics of the membrane from which one may obtain the membrane
roughness. Since our main goal in this work is to study the
behavior of lipid membranes in the presence of an external
electric field, it would be advantageous to write the above
Hamiltonian when such a field is present. Therefore, in what
follows, we study a membrane under the influence of an electric
field and investigate the resulting effects that such a field may
have, using a modified form of the above Hamiltonian.
\subsection{The effective surface tension in an electric field}
Since lipid bilayers are impermeable to ions, in the presence of an
external electric field, charges accumulate at the bilayer
interface, as shown in Fig. 1. Thus, an additional transmembrane
potential is created across the membrane. An almost flat, infinite
and poorly conductive membrane under the influence of a fixed
external electric field may be used as a simple model to investigate
the effects of the electric field on lipid membranes, see Fig. 1.

One may embark on obtaining a relation between the local undulation
of lipid membranes in the large-wavelength limit by using the
discontinuity of the Maxwell stress tensor across the membrane
interface and solving the corresponding electro-kinetic problem by
employing the usual boundary conditions on displacement vector and
current density. This method was employed in \cite{isam1}, leading
to a net decrease in energy at long wavelength limit relative to the
membrane thickness $d$. This extra electric field contribution to
the Canham-Helfrich Hamiltonian can be written as
\begin{eqnarray}
{\cal H}_{el}=\frac{\sigma_{el}}{2}\int d^2{x}(\nabla h)^2~.
\label{eqn1}
\end{eqnarray}
Here $\sigma_{el}$ is an induced \emph{effective negative surface
tension}
\begin{eqnarray}
\sigma_{el}=-\alpha E^2, \label{eqn03}
\end{eqnarray}
where $$\alpha=d~\epsilon_m\left[\frac{g_1}{g_m}\right]^2,$$ and $E$
is the electric field strength. As for the other quantity
$\epsilon_m$ refers to the permittivity of the membrane and $g_m$,
$g_1$ are the conductivities of the lipid membrane and the
surrounding medium respectively.

\begin{figure}
\centering
\includegraphics[width=8.25cm]{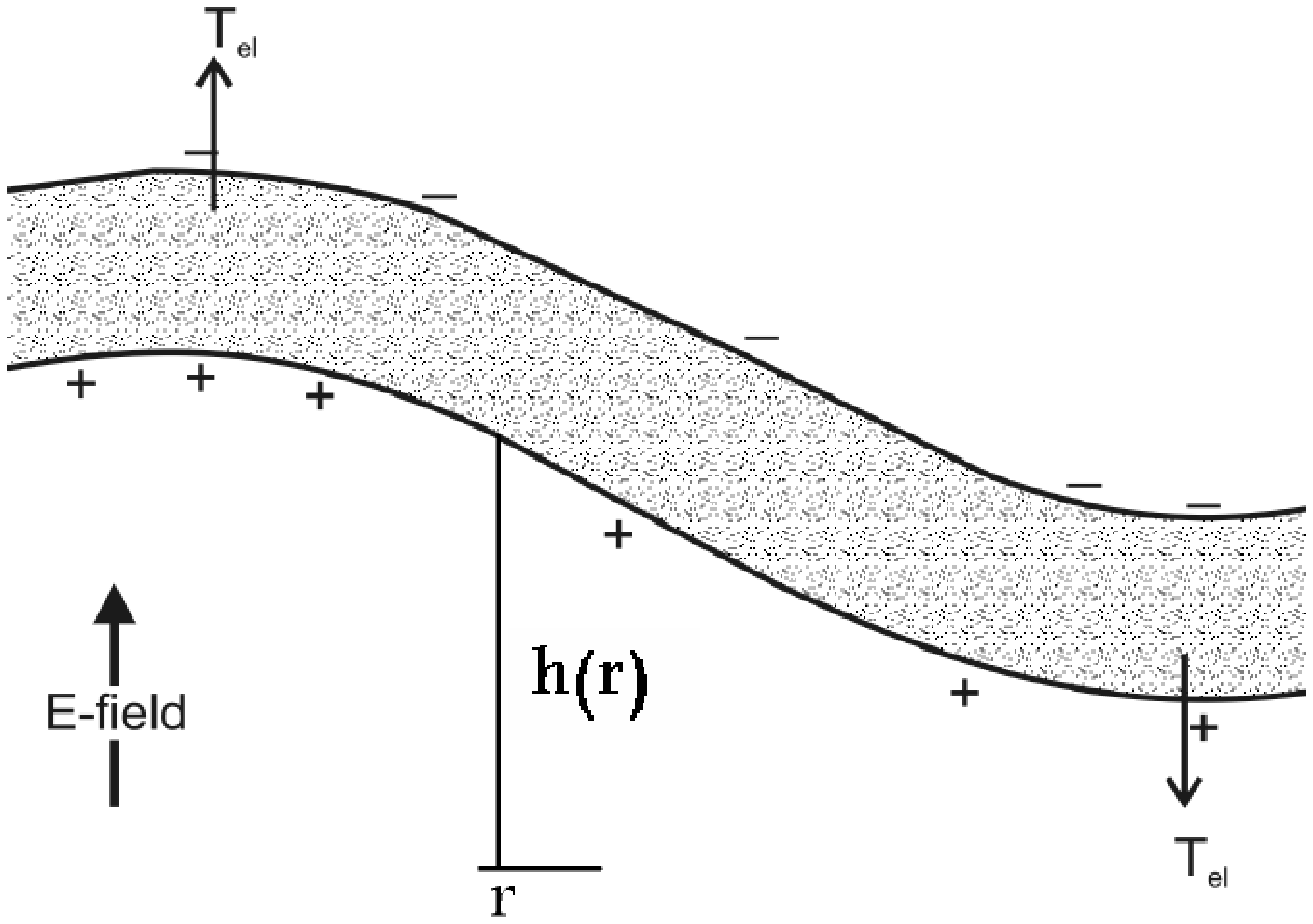}
\caption{Net accumulation of conduction charges near a curved lipid
membrane in an electric field, figure is based on ref \cite{isam1}.
T$_{el}$ is the resulting normal stress component in the
$z$-direction which enhances the local undulations h(\textbf{r}) of
the membrane. It also induce a further surface tension in the
membrane surface.} \label{fig:bz}
\end{figure}

\begin{figure}
\centering
\includegraphics[width=8.25cm]{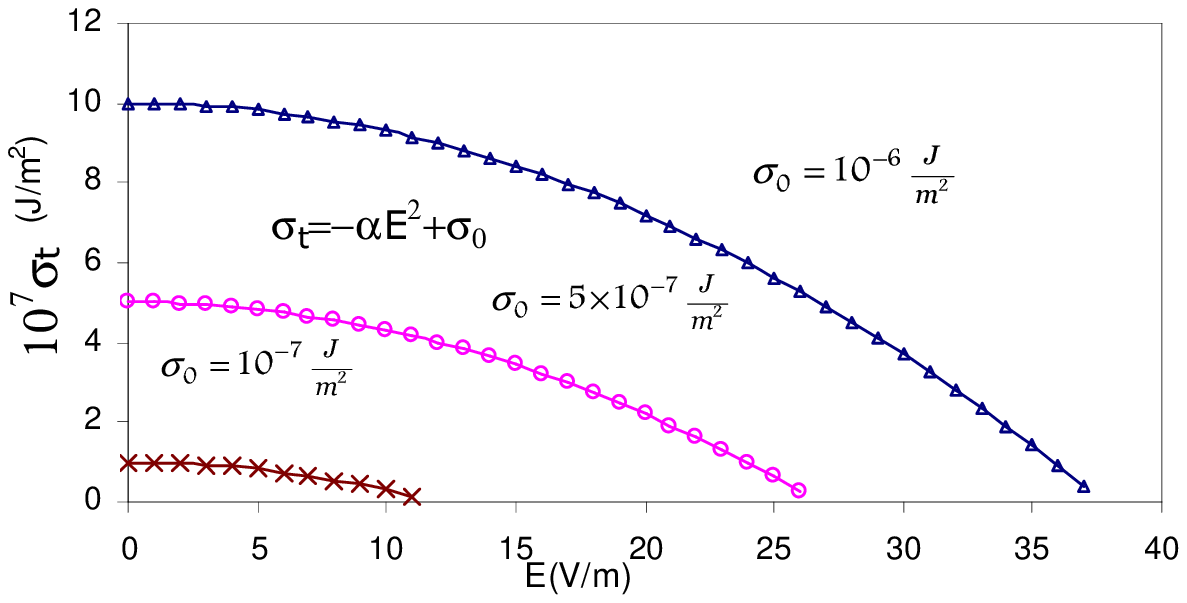}
\caption{Total surface tension versus the electric field for three
different values of $\sigma_0$.} \label{fig:bz}
\end{figure}

The similarity of the equation (\ref{eqn1}) and the second term on
the RHS of equation (\ref{eq1}) is clear. Both represent the surface
tension contributions to the Hamiltonian, one relating to the
cytoskeleton-generated surface tension and the other only appears
when an external electric field is applied.

For typical values of the parameters quoted in \cite{isam1}, one
gets $\alpha=0.71\times10^{-9}$ $\frac{J}{V^2}$. Therefore, the
total membrane surface tension $\sigma_t$ (which appear in the
surface tension part of the Hamiltonian) as the sum of electricaly
induced part $\sigma_{el}$ and cytoskeleton-generated membrane
surface tension $\sigma_0$ may be written as
\begin{eqnarray}
\sigma_t=\sigma_{el}+\sigma_0. \label{eqn33}
\end{eqnarray}
The electric part of surface tension $\sigma_{el}$, being
negative, enhances the undulations of the membrane while
$\sigma_0$, the intramembrane tension, tends to keep the membrane
surface unaltered. Note that we use positive values for
$\sigma_{t}$ to satisfy the nearly-flat assumption. Typical values
for the intramembrane tension and bending rigidity of a
fluctuating membrane at room temperature for different lipid
materials are $10^{-9} \frac{J}{m^2}\leq\sigma_0\leq
10^{-6}\frac{J}{m^2}$ \cite{Raf2} and $\kappa_0\sim 10^{-20}J$
($\simeq2.5k_BT$) \cite{kummorow}, respectively. These values
correspond to the following range of values for the electric
field, using equation (\ref{eqn03}) and (\ref{eqn33}), presented
in Table I.
\begin{center}
\begin{tabular}{cc}
$\sigma_0(\frac{J}{m^2})$&$E(\frac{V}{m})$\\
\hline \hline\\
$10^{-6}$&$37.9$\\
$10^{-7}$&$12.0$\\
$10^{-8}$&$3.5$\\
$10^{-9}$&$1.2$\\
\hline
\end{tabular}
\end{center}
\begin{center}{\footnotesize Table I: Upper bounds for the electric field
corresponding to arbitrary intramembrane surface tension
$\sigma_0$, used in the present work. As can be seen, the weaker
the electric field, the lower the value of $\sigma_0$. A membrane
with larger $\sigma_0$ can withstand higher electric fields. }
\end{center}
The above values for the electric field have been taken with the
view that the assumption of a nearly flat membrane is satisfied and
is not violated by the application of too strong an electric field.
They indicate that a membrane with weak surface tension is more
affected by an electric field than a membrane with a stronger
surface tension. Figure 2 shows variations of $\sigma_t$ with
respect to the electric field for three values of $\sigma_0$.

\subsection{The effective bending rigidity in an electric field}
One can assume a similar equation for the total bending rigidity
$\kappa_t$ as the sum of the electric contribution $\kappa_{el}$ and
the cytoskeleton-generated rigidity $\kappa_0$, that is
\begin{eqnarray}
\kappa_t=\kappa_{el}+\kappa_0. \label{eqn4}
\end{eqnarray}
In what follows, we will make an estimate of $\kappa_{el}$ using
the theory of charged membranes and equation (\ref{eqn33}) and
show that $\kappa_{el}$ is two order of magnitude smaller than
$\kappa_0$ used below.

Let us assume that the membrane is charged due to the influence of
an external electric field and the resulting accumulated net charge
fluctuates on the surface. In such a scenario the contributions of
$\sigma_{el}$ and $k_{el}$ to the surface tension and bending
rigidity which  stem from charge fluctuations, are both negative. In
the limit of large membrane undulations, the following relation for
the negative surface tension may be written \cite{ywkim}
\begin{eqnarray}
\sigma_{el}=-\frac{\pi}{2\lambda^2}k_BT~\left[\frac{2\pi\lambda}{a}-
\ln\left(1+\frac{2\pi\lambda}{a}\right)\right],\label{eqn303}
\end{eqnarray}
where $\lambda$ is a characteristic length similar to that of the
Gouy-Chapman length especially in the presence of counterions
around the membrane and $a$ is the molecular cut-off. The solution
of the Poisson-Boltzmann equation for a simple charged surface in
the presence of counterions would enable one to define a
characteristic length which decreases when the surface charge
density goes to higher values. In the present study the membrane
is charged via the external electric field and is not flat.
However, we have assumed that the situation is similar to the case
of a charged surface when $\lambda\gg a$. In this limit, we may
safely ignore the logarithm term in equation (\ref{eqn303})
against $2\pi\lambda/a$. Equations (\ref{eqn03}) and
(\ref{eqn303}) then implicitly lead to an expression for the
characteristic length, $\lambda$, mentioned earlier
\begin{eqnarray}
\lambda\simeq \frac {k_BT}{\epsilon_m}{\left(\frac{\pi g_m}{g E
d}\right)}^2. \label{eqn44}
\end{eqnarray}
As can be seen, an increasing electric field causes $\lambda$ to
decrease as a result of the increasing charge density, leading to a
highly charged membrane. Also, when $a\rightarrow 0$, one has
$\lambda\rightarrow \infty$, showing that for a small molecular
cut-off, the membrane becomes poorly charged and that in the
presence of counterions, the charges do not accumulate near the
membrane and often move freely in the solvent. On the other hand, if
$\frac{g_m}{g}$ decreases or $\kappa_m$ increases, $\lambda$
increases, since there would be a concentration of charges on the
boundaries. The direct dependence of $\lambda$ on temperature is
clearly seen. The dependence of $\lambda$ on the electric field
strength and membrane conductivity is shown in Fig. 3.

In the same manner, if we assume that the  membrane is charged by
the application of the electric field so that net charges on the
membrane can fluctuate, in the long wavelength limit, the following
expression for the reduced bending rigidity holds \cite{ywkim}
\begin{eqnarray}
\kappa_{el}=-\frac{\pi}{32}k_B
T\ln\left(\frac{L+2\pi\lambda}{a+2\pi\lambda}\right), \label{eqn45}
\end{eqnarray}
where $L$ is a typical membrane length taken as $40 \,\mu\mbox{m}$
here. Figure 4 shows variation of $\kappa_{el}$ with the electric
field and $g_m$ for $g=1 \frac{S}{m}$, $a=2 nm$, $T=300 K$ and
$\epsilon_m=2\epsilon_0$. Note that the membranes become more
flexible for higher values of both the field strength and membrane
conductivity. This flexibility decreases rapidly when the membrane
conductivity is low and the electric field strength is weak. It
should be emphasized that the theory explained above may also be
used when other sources are present. It has been used to study
membranes with excess surface charge density \cite{ywkim} by
dividing the Hamiltonian into three terms, that is, electrostatic
interaction, entropic contribution and elastic energy. The effects
of ionic salts on bending rigidity around cell membranes and in
particular of micron-sized vesicles are an interesting problem to
look at in that the combined effects of the salt and electric field
induce cylindrical deformations \cite{dimova} on such vesicles.

\begin{figure}
\centering
\includegraphics[width=8.25cm]{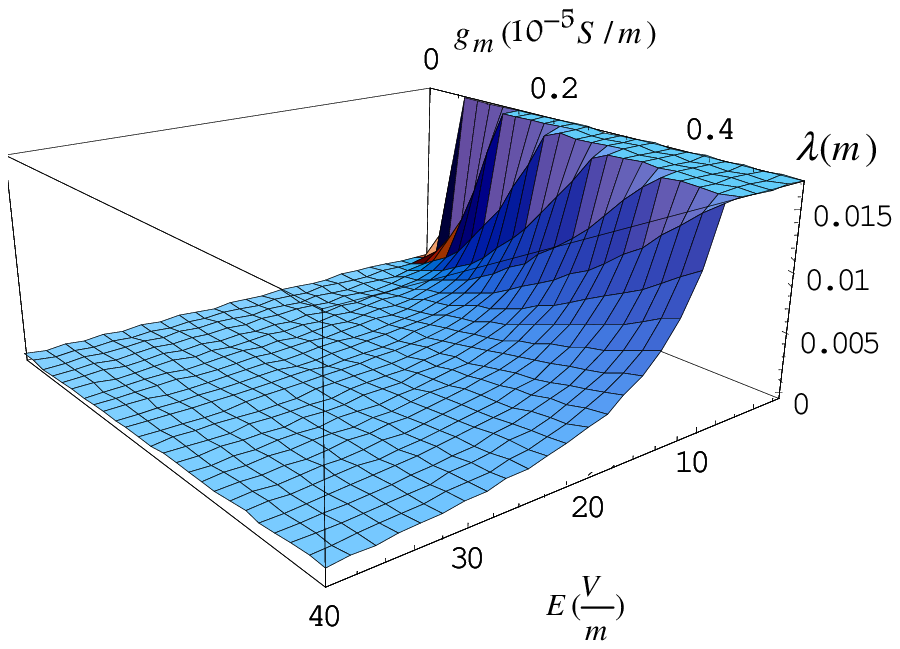}
\caption{Variation of $\lambda$ in terms of different values of $E$
and $g_m$ for $g=1 \frac{S}{m}$.} \label{fig:bz1}
\end{figure}

\begin{figure}
\centering
\includegraphics[width=8.25cm]{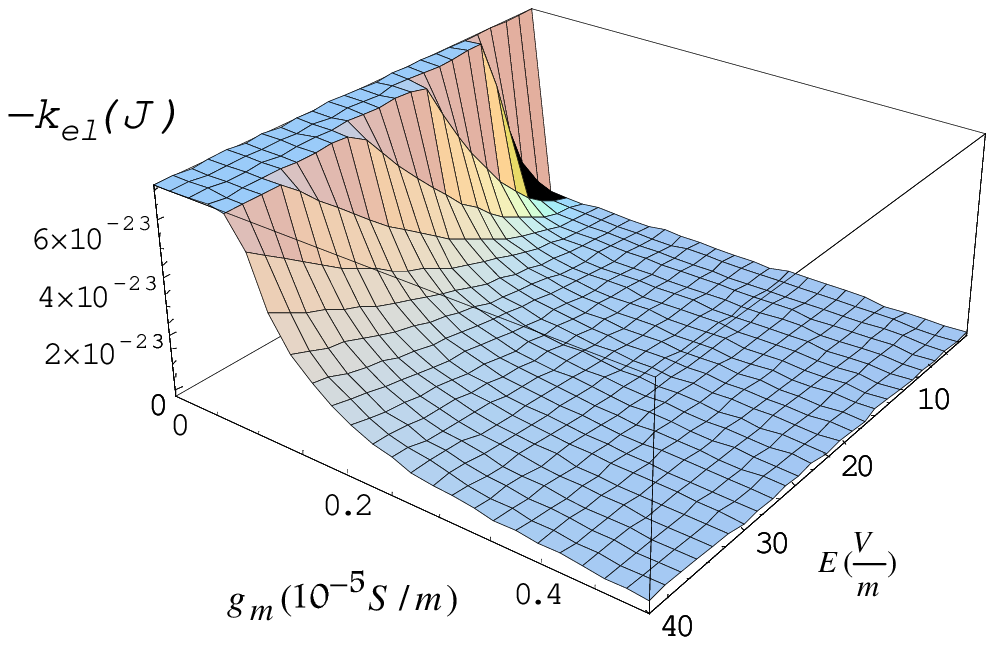}
\caption{Variation of $-\kappa_{el}$ in terms of different values of
$E$ and $g_m$ for $g=1 \frac{S}{m}$.} \label{fig:bz2}
\end{figure}

The above discussions leads us, using equations (\ref{eqn33}) and
(\ref{eqn4}), to write the Hamiltonian as
\begin{eqnarray}
{\cal H}=\frac{1}{2}\int d^2x\left\{\kappa_t(\nabla^2
h)^2+\sigma_t(\nabla h)^2\right\}. \label{eq3}
\end{eqnarray}
This form is particularly useful for deriving the height-height
correlation function, from which the roughness associated with the
membrane can be deduced.
\section{The influence of an electric field on the roughness of lipid membranes}
The study of roughness in lipid membranes is facilitated by treating
the evolution of such a system in terms of stochastic processes. One
way to introduce stochastic behavior into the dynamics of the
membrane is to treat the height function, $h(x_1,x_2)$, as a
stochastically fluctuating Wiener variable. This stochastic behavior
can be communicated to the inclusions residing inside, and on the
surface of the membrane. One may then proceed to obtain the
height-height correlation function of the membrane in real space
from which the \emph{roughness}, associated with random changes in
$h$, can be extracted. In the presence of an electric field and
using the the Fourier representation of $h$, we arrived at the
Hamiltonian \cite{Raf2}
\begin{eqnarray}
{\cal H}=\frac{1}{2}\int\frac{d^2 q}{(2\pi)^2}\left\{\kappa_t q^4+
\sigma_t  q^2\right\}h({\bf q})h^\ast({\bf q})\,, \label{eq5}
\end{eqnarray}

\begin{figure}
\centering
\includegraphics[width=7cm]{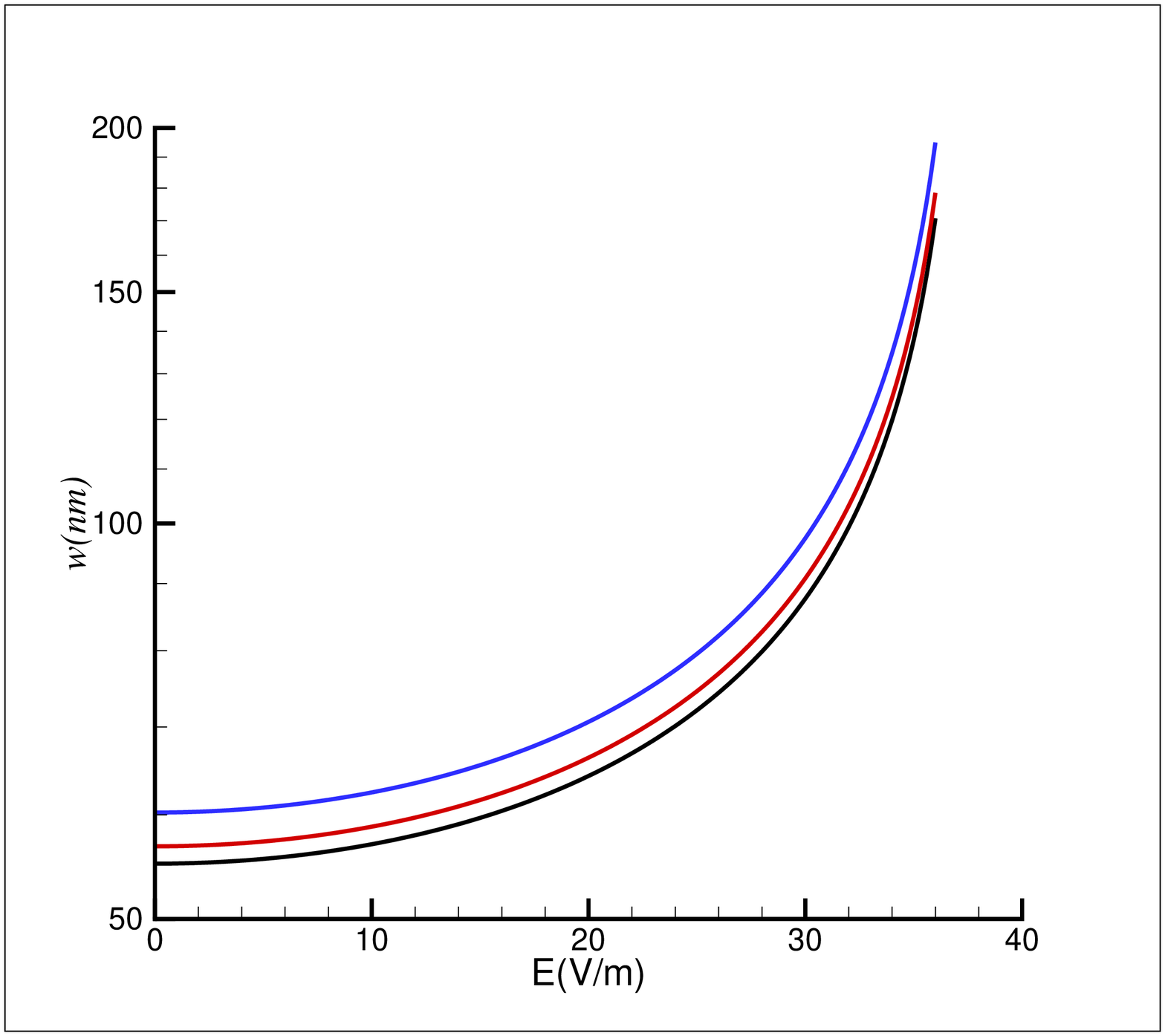}
(a)
\includegraphics[width=7cm]{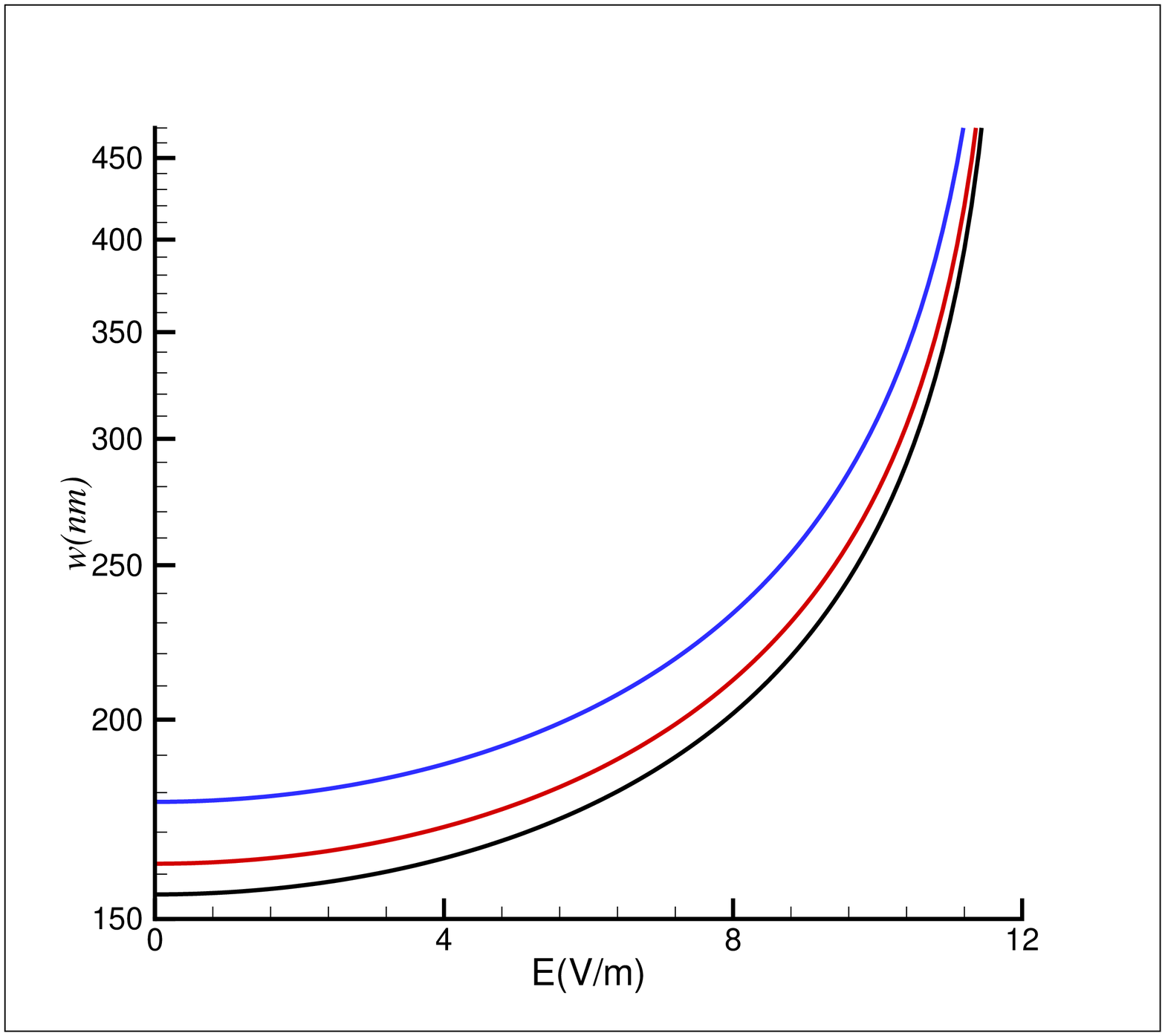}
(b)
 \caption{Variation of the lipid membrane's \emph{roughness}
versus the electric field for $L=40\mu m$, where
$\sigma_0=10^{-6}\frac{J}{m^2}$ in the (a) panel and
$\sigma_0=10^{-7}\frac{J}{m^2}$ in the (b) panel. In each panel
$\kappa_0=10^{-20}(J),5\times10^{-20}(J)$ and $ 10^{-19}(J)$
correspond to the top (blue), middle (red) and bottom (black)
curves.
 } \label{fig:bz1}
\end{figure}

\begin{figure}
\centering
\includegraphics[width=7cm]{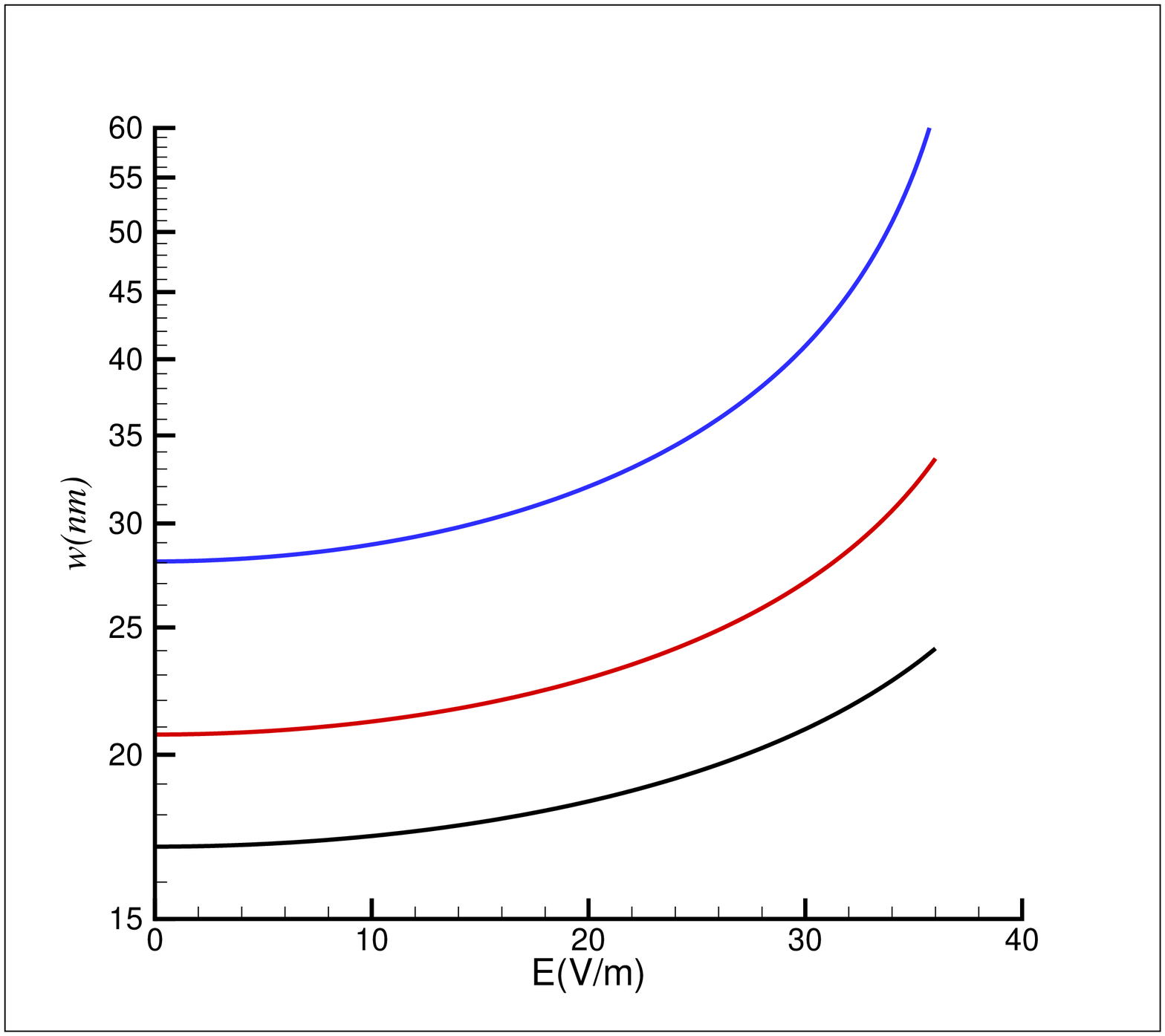}
(a)
\includegraphics[width=7cm]{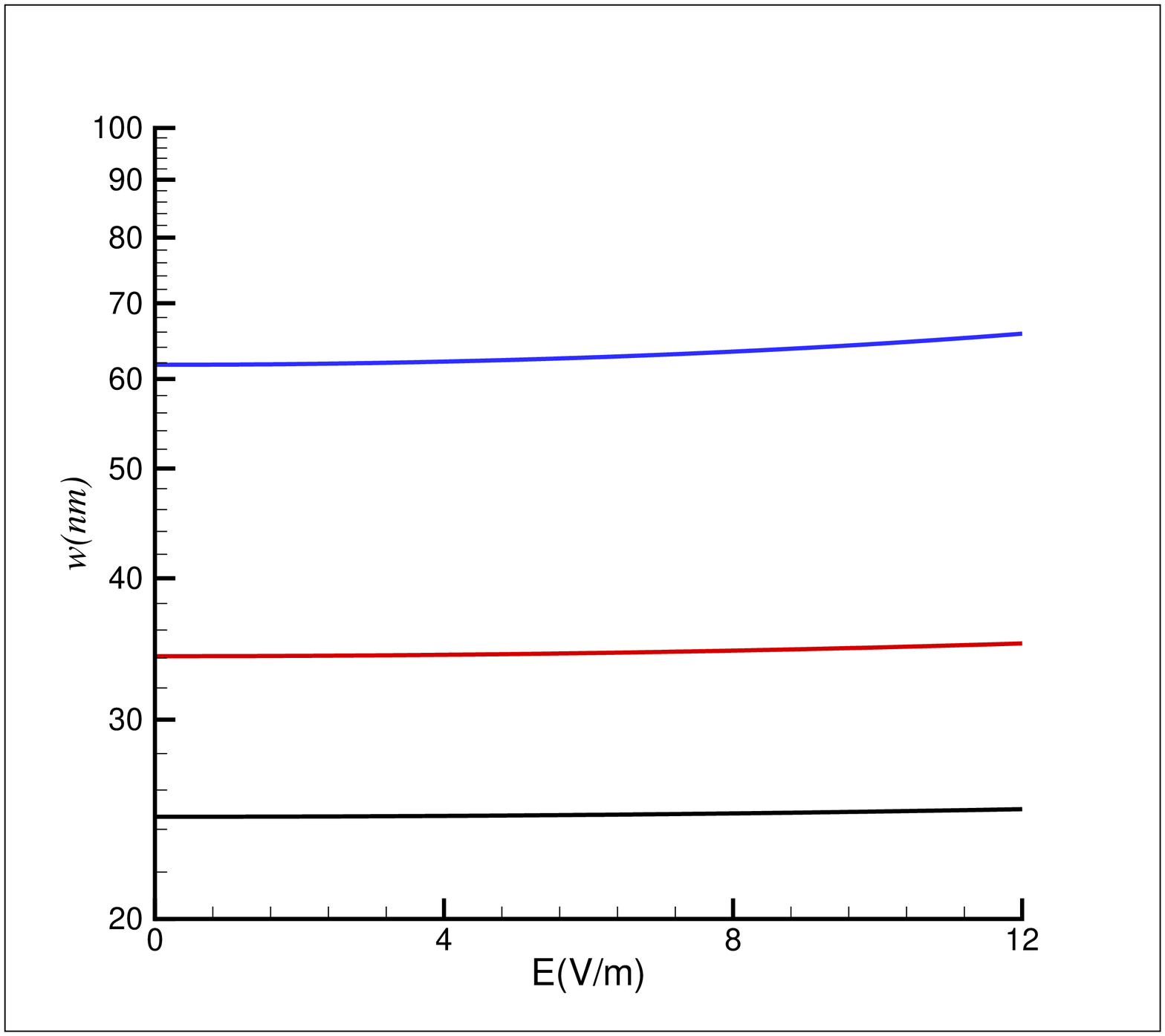}
(b)
 \caption{Variation of the lipid membrane's \emph{roughness}
versus the electric field for $L=500 nm$, where
$\sigma_0=10^{-6}\frac{J}{m^2}$ in the (a) panel and
$\sigma_0=10^{-7}\frac{J}{m^2}$ in the (b) panel. In each panel
$\kappa_0=10^{-20}(J),5\times10^{-20}(J)$ and $ 10^{-19}(J)$
correspond to the top (blue), middle (red) and bottom (black)
curves.
 } \label{fig:bz2}
\end{figure}
where $*$ denotes the complex conjugate. The {\it static}
height-height correlation function can then be calculated as follows
\begin{eqnarray}
\langle h({\bf q};0)h^\ast({\bf
q^\prime};0)\rangle=\left(\frac{k_{\tiny\mbox{B}} T } {\kappa_t
q^4+\sigma_t q^2}\right)(2\pi)^2\delta({\bf q}-{\bf q}^\prime).
\label{eq6}
\end{eqnarray}
The averaging is done with respect to the Boltzmann weight factor
$\mbox{exp}(-\frac{\cal H}{k_{\tiny\mbox{B}} T})$, with
$k_{\tiny\mbox{B}}$ being the Boltzmann constant. One can write
the {\it dynamic} correlation function as \cite{Seifert1}
\begin{eqnarray}
h({\bf q};t)=h({\bf q};0)e^{-\gamma(q)t},\nonumber
\end{eqnarray}
leading to
\begin{eqnarray}
\langle h({\bf q};t)h^\ast({\bf q^\prime};0)\rangle=
 \left(\frac{k_B T~e^{-\gamma(q)t}} {\kappa_t q^4+\sigma_t
q^2}\right)(2\pi)^2\delta({\bf q}-{\bf q}^\prime), \label{eq7}
\end{eqnarray}
where $\gamma(q)$ is the damping factor reflecting the long-range
character of the hydrodynamic damping given by
\begin{eqnarray}
\gamma(q)=\frac{1}{4\eta}\left(\kappa_t q^3+\sigma_t q\right).
\label{eq8}
\end{eqnarray}
Let us now suppose that the fluid surrounding the membrane has a
viscosity $\eta$. It is then possible to write  the Fourier
transform of (\ref{eq7}) in real space (i.e, $\langle h({\bf
x};t)h({\bf x};0)\rangle$), representing the dynamical mean
\emph{roughness} of the membrane as
  \begin{eqnarray}
w^2=\int\int\frac{d^2 q}{(2\pi)^2}\frac{d^2
q^{\prime}}{(2\pi)^2}\langle h({\bf q};t)h^*({\bf
q^{\prime}};0)\rangle e^{-{\bf x}.({\bf q}-{\bf
q^{\prime})}},\label{eq8a}
\end{eqnarray}
Substituting from (\ref{eq7}) and (\ref{eq8}) we find
\begin{equation}
w^2=\frac{1}{2\pi\beta}\int_{\frac{1}{L}}^{\frac{1}{a}}dq
\frac{e^{-\frac{1}{4\eta}(\kappa_t q^3+\sigma_tq)t}}{\kappa_t
q^3+\sigma_t q}, \label{eq9}
\end{equation}
where $L$ is the linear size of the membrane and $a$ is the
molecular cut-off, of the order of a nanometer, and
$\beta=\frac{1}{k_{\tiny\mbox{B}}T}$. Here $t$ represents the
correlation time \cite{Seifert1} and is of the order of $10^{-4}s$
for the said values. Since there is no closed form expression for
the integral given in equation (\ref{eq9}), it should be computed by
expanding the integrand as a series. The square root of $w^2$ is
called root-mean-squared roughness (rms) when one assumes that
$\langle h(x,t)\rangle=0$. Figure~5 and Fig.~6 demonstrate the
variation of $w$  with respect to an electric field. We have limited
the electric field values to those presented in Table I. Here, the
three different values for the bending rigidity
$\kappa_0=10^{-20}(J), 5\times10^{-20}(J)$ and $ 10^{-19}(J)$
correspond to the top, middle and bottom curves in each panel
respectively. Note that in the (a) panels,
$\sigma_0=10^{-6}\frac{J}{m^2}$ while in the (b) panels
$\sigma_0=10^{-7}\frac{J}{m^2}$. Also the typical values have been
used as $a=2nm$, $T=300K$, and $\eta=10^{-3} \frac{Js}{m^3}$.

Note that the dimension of roughness in the figures is nanometer.
The following observations worth mentioning: that the roughness
values are of the order of a few hundred nanometers and that the
decrease in $\kappa_0$ and $\sigma_0$ and the increase in $L$ cause
the roughness to increase. Here, it would be useful to mention that
the observations of a typical membrane with $L=10\,\mu\mbox{m}$ by
AFM have yielded a roughness in the range 12-70 nanometers
\cite{langmuir}. Also, results from molecular dynamics and monte
carlo simulations of lipid membranes with sizes of the order of
nanometers \cite{jchemphys} show that they experience a roughness of
the order of a few angstroms. For stronger fields, we see a dramatic
change in the roughness, starting roughly at $30 \frac{V}{m}$ for
$\sigma_0=10^{-6}\frac{J}{m^2}$ and $10 \frac{V}{m}$ for
$\sigma_0=10^{-7}\frac{J}{m^2}$. Therefore, the presence of an
external electric field leads to the undulations or roughening of
the membrane surface. Smaller values for correlation time lead to a
lesser increase in rms \emph{roughness} and vice versa.
\section{Conclusions}
In this paper we have studied the behavior of lipid membranes in the
presence of an external electric field. The negative contribution to
the surface tension as a result of the application of the field, the
effects of an external electric field on the roughening of membranes
and an estimation of the thickness of charges aggregated on the
membrane surface were reviewed. The dependence of bending rigidity
on electric fields and its relation to the thickness mentioned above
was studied. The increase in the membrane rms roughness for larger
electric field values was calculated and shown to contribute to the
roughening of the membrane surface.

\end{document}